\documentclass[aps,reprint,prd,superscriptaddress,nofootinbib,amsmath,amssymb]{revtex4-2}

\usepackage{xcolor}
\usepackage{mathrsfs}
\usepackage[colorlinks=true,citecolor=blue]{hyperref}
\usepackage{orcidlink}

\def\rd{{\rm d}}

\begin{document}

\title{The dyonic Kerr-Schild ansatz}

\author{Eloy \surname{Ay\'{o}n-Beato}\,\orcidlink{0000-0002-4498-3147}}
\email{eloy.ayon-beato@cinvestav.mx}
\affiliation{Departamento de F\'{i}sica, Cinvestav, Av.~IPN 2508, 07360, CDMX, M\'exico}

\author{Daniel \surname{Flores-Alfonso}\,\orcidlink{0000-0001-7866-3531}}
\email{dafa@azc.uam.mx}
\affiliation{Departamento de Ciencias B\'asicas, Universidad Aut\'onoma Metropolitana -- Azcapotzalco, Avenida San Pablo 420, Colonia Nueva El Rosario, Azcapotzalco 02128, Ciudad de M\'exico, Mexico}

\author{Mokhtar \surname{Hassaine}\,\orcidlink{0009-0003-3159-5916}}
\email{hassaine@inst-mat.utalca.cl}
\affiliation{Instituto de Matem\'{a}ticas (INSTMAT), Universidad de Talca, Casilla 747, Talca 3460000, Chile}

\author{Daniel F.~\surname{Higuita-Borja}\,\orcidlink{0000-0001-5630-4343}}
\email{dfhiguit@gmail.com}
\affiliation{Instituto de F\'isica, Benem\'erita Universidad Aut\'onoma de Puebla, Edificio IF-1, Ciudad Universitaria, Puebla, Puebla 72570, M\'exico}
\affiliation{Instituto de F\'isica, Universidad de Antioquia, Calle 70 No 52-21, Medell\'in, Colombia}

\begin{abstract}
We develop a geometric extension of the Kerr–Schild ansatz that incorporates both electric and magnetic sectors of the Maxwell field in a unified framework, without resorting to duality rotations. We start observing that the known purely electric solution satisfies Maxwell’s equations due to a closedness condition obeyed by the Kerr–Schild null congruence. From the associated local exactness property, we construct a new one-form naturally linked to the congruence as a sort of Poincar\'e dualization. This leads us to propose a geometrically motivated dyonic vector potential within the Kerr–Schild ansatz, defined as a superposition of an electric contribution along the congruence and a magnetic one that aligns to the dualized one-form. We then show that for a stationary and axisymmetric Kerr–Schild ansatz, the electrovac circularity theorem uniquely constrains not only the scalar profile of the metric, but also those associated to the electric–magnetic splitting of the gauge field. The resulting formalism provides a transparent derivation of the dyonic Kerr–Newman solution and extends naturally to the (A)dS case, highlighting the intrinsic interplay between geometry and matter in a Kerr–Schild setting.
\end{abstract}

\maketitle

%%%%%%%%%%%%%%%%%%%%%%%
\section{Introduction}
%%%%%%%%%%%%%%%%%%%%%%%

The Kerr–Schild ansatz \cite{Kerr:1965} has long stood as one of the most elegant and effective techniques for generating exact solutions to Einstein’s field equations. In this setting, the spacetime metric $g$ is expressed as a linear deformation of a seed background $g^{(0)}$ by a term quadratic in a shear-free and geodesic null vector $l$
\begin{equation}\label{eq:KSansatz}
g = g^{(0)} + 2S\,l \otimes l,
\end{equation}
modulated by a scalar profile $S$. This linearization allows the highly nonlinear Einstein equations to be reduced to more tractable linear differential equations for the profile. The class of spacetimes admitting a Kerr–Schild representation includes a wide range of physically significant solutions. Outstandingly, this ansatz precisely selects the black hole configurations within the whole stationary solutions: such as the Schwarzschild–(A)dS black holes and their spinning Kerr–(A)dS counterparts in vacuum \cite{Kerr:1963ud,Carter:1968ks,Carter:1972,*Carter:2009nex,Plebanski:1975xfb}, as well as their four-dimensional Kerr–Newman charged versions in electrovac \cite{Newman:1965my,Demianski:1966}. In the time-dependent context it also describes classes of exact radiative spacetimes such as \emph{pp}-waves and AdS-waves, see e.g.~\cite{Ayon-Beato:2005gdo,Ayon-Beato:2018hxz}.

The Kerr–Schild formalism builds on the Kerr theorem \cite{Debney:1969zz,Cox:1976,Stephani:2003tm}, which classifies shear-free and geodesic null congruences in flat spacetime. However, its original formulation is strongly degenerate, yielding infinitely many congruences unless further conditions are imposed. A way to circumvent this issue is through a symmetry-based refinement as proposed in \cite{Ayon-Beato:2015nvz}, where imposing stationarity and axisymmetry uniquely selects the congruence leading to the Kerr vacuum solution \cite{Kerr:1965}. This idea was further pursued in \cite{Ayon-Beato:2025owr} by considering an ultrarelativistic limit, requiring invariance under null translations and axisymmetry. In doing so, only two admissible congruences are obtained, both associated with physically relevant solutions.

Beyond vacuum configurations, the Kerr–Schild ansatz has proven remarkably effective in coupling gravity to a Maxwell field, where the vector potential is assumed to be aligned with the Kerr–Schild congruence \cite{Ayon-Beato:2015nvz,Ayon-Beato:2015qtt}
\begin{equation}\label{eq:electric}
    A = -S_\text{e} \, l.
\end{equation}
Here $S_\text{e}$ is a second scalar profile, and when the whole ansatz is inserted into the Einstein-Maxwell equations the system remains analytically solvable, leading to celebrated solutions such as the electrically charged Kerr–Newman black hole \cite{Newman:1965my}. This reflects a profound compatibility between the structure of gauge fields and the geometry encoded in the Kerr–Schild formalism.

However, the described approach is incomplete since the Kerr–Newman black hole also exhibits a magnetic contribution \cite{Demianski:1966} which is lacking in the classical Kerr-Schild method. The common way to circumvent this issue is to appeal to an important aspect of the Einstein–Maxwell equations of motion, namely, the symmetry of the electromagnetic field under electric–magnetic duality rotations, see Ref.~\cite{Ayon-Beato:2024vph} for a recent account. This symmetry allows one to map purely electric solutions to their magnetic or dyonic counterparts, and has often been used as a convenient tool to generate new families of charged spacetimes from known ones. In particular, magnetically charged black holes are frequently obtained by applying such duality operations to electrically charged configurations, thereby bypassing the need to solve the coupled field equations anew. However, this strategy tends to obscure the role of magnetic and dyonic components at the level of the original ansatz. From a geometric and physical standpoint it is desirable to construct these solutions directly, without appealing to duality, by incorporating the magnetic sector within the foundational structure of the Kerr–Schild framework itself. This requires a more general treatment of the gauge potential and field strength, one that accommodates both electric and magnetic contributions in a unified manner. Such an approach not only enhances the transparency of the underlying geometry-matter coupling, but also opens the door to discovering genuinely new solutions that may elude traditional duality-based constructions.

In this work, we present a novel generalization of the Kerr–Schild ansatz that achieves a natural and unified incorporation of both electric and magnetic components of the gauge field. The central innovation is a dualization procedure applied directly within the Kerr–Schild framework, which allows us to construct the magnetic sector not by invoking external duality arguments, but through a purely geometric extension of the ansatz itself. Starting from the standard electrically charged stationary axisymmetric configuration \cite{Ayon-Beato:2015nvz} in which the electric vector potential is aligned with the shear-free and geodesic null congruence, $A_\text{e}\propto l$, we show that the null tangent vector provides sufficient information to construct a new one-form denoted by $l^{*}$ through a procedure reminiscent of Poincar\'e dualization \cite{Ortin:2004ms}. More precisely, this construction is based on a direct application of the Poincar\'e lemma to the Maxwell electric equations, and directly leads to the one-form field $l^{*}$ that naturally captures the magnetic part of the vector potential, namely $A_\text{m}\propto l^{*}$. Remaining in the Kerr-Schild framework, we then consider a superposition of both contributions for the gauge potential, $A=A_\text{e}+A_\text{m}$. Remarkably, the electrovac circularity theorem applicable to this larger stationary axisymmetric configuration, uniquely constrains each proportionality profile up to gauge transformations. The result is a genuinely dyonic extension of the Kerr–Schild ansatz, in which both electric and magnetic sectors emerge from a purely geometric and non-trivial construction, without relying on external duality transformations. Far from being a subsequent input, the magnetic part is shown to emerge as a natural companion to the electric sector, dictated by geometry itself.

This generalized construction leads to a transparent and fully self-contained derivation of the dyonic Kerr–Newman black hole \cite{Demianski:1966}, entirely accomplished within the Kerr–Schild formalism. The resulting solution captures both electric and magnetic charges in a symmetric and covariant manner, reaffirming the deep geometric compatibility between the gauge field structure and the null congruence that defines the Kerr–Schild metric. Moreover, we establish the versatility of this extended framework by extending its applicability in the presence of a cosmological constant. By incorporating a constant curvature background as the seed metric of the formalism, we recover the dyonic Kerr–Newman–(A)dS black hole \cite{Carter:1968ks,Carter:1972,*Carter:2009nex,Plebanski:1975xfb} within the same dyonic Kerr–Schild setting. 

The plan of the paper is as follows. In the next section, the derivation of the purely electric Kerr-Newman black hole within the stationary and axisymmetric version of the Kerr theorem \cite{Ayon-Beato:2015nvz} is reviewed in order to be self-contained. In Sec.~\ref{sec:dyonic}, we will show how starting from the shear-free and geodesic null congruence above that defines the electrically charged Kerr–Newman solution, one is naturally led to construct through a dualization procedure a one-form that proves useful in capturing the magnetic sector of the vector potential. Using a new ansatz incorporating both contributions for the Maxwell gauge field, we will demonstrate how the electrovac circularity theorem uniquely constrains both the electric and magnetic profiles, thereby yielding the dyonic solution in a purely geometric manner. Sec.~\ref{sec:cosmological} is devoted to exhibit that this construction can be naturally extended in presence of a cosmological constant. The final section presents our conclusions.

%%%%%%%%%%%%%%%%%%%%%%%%%%%%%%%%%%%%%%%%%%%%%%%%%%%%%%%%%
\section{Kerr-Newman from Kerr-Schild: Purely Electric Case}
%%%%%%%%%%%%%%%%%%%%%%%%%%%%%%%%%%%%%%%%%%%%%%%%%%%%%%%%%

The electrically charged Kerr-Newman solution \cite{Newman:1965my} was derived in \cite{Ayon-Beato:2015nvz} by extending a symmetric refinement of the Kerr theorem. That approach proves highly useful for the dyonic extension. Thus, we briefly review the main steps here in order to be as self-contained as possible. First of all, it was established in \cite{Ayon-Beato:2015nvz} that the tangent vector to the unique shear-free and geodesic null congruence on flat spacetime invariant under stationarity and axisymmetry is naturally expressed in ellipsoidal coordinates as
\begin{equation}\label{eq:lEllips}
l=\rd t - a\sin^2{\theta}\rd\phi 
+ \frac{\Sigma}{r^2+a^2}\rd r,
\end{equation}
where the Minkowski seed metric turns out to be 
\begin{equation}\label{eq:FlatEllips}
\rd s_{0}^2= -\rd t^2
+(r^2+a^2)\sin^2{\theta}\rd\phi^2 +\frac{\Sigma}{r^2+a^2}\rd r^2
+\Sigma \rd\theta^2,
\end{equation}
with $\Sigma=r^2+a^2\cos^2\theta$. Since we are interested in a stationary and axisymmetric Kerr-Schild ansatz, both for the metric \eqref{eq:KSansatz} and the vector potential \eqref{eq:electric}, the corresponding  profiles are restricted to the dependence $S=S(r,\theta)$ and $S_\text{e}=S_\text{e}(r,\theta)$.

Now, let us recall that every stationary axisymmetric spacetime admits two commuting Killing vector fields, say $k=\partial_t$ and $m=\partial_\phi$~\cite{Carter:1970ea}. Moreover, such vectors satisfy a pair of originally geometrical identities which constitute the basis of the circularity theorem \cite{Papapetrou:1966zz,Kundt:1966zz,Carter:1969zz}. For Einstein-Maxwell theory these identities become \cite{Heusler:1996}
\begin{equation}\label{eq:MaxId}
\begin{aligned}
    -\frac{1}{4}\rd \star (k\wedge m \wedge \rd k) &= \star F(k,m)E_k + F(k,m)B_k, \\
    -\frac{1}{4}\rd \star (k\wedge m \wedge \rd m) &= \star F(k,m)E_m + F(k,m)B_m,
\end{aligned}
\end{equation}
where for each Killing vector field $X$ we have defined the electric and magnetic one-forms $E_X=-\iota_X F$ and $B_X=\iota_X\star F$, respectively. On the other hand, for stationary-axisymmetric electromagnetic fields, $\pounds_k F = 0 = \pounds_m F$, Maxwell's equations ultimately imply that the following smooth components are global constant vanishing at the symmetry axis where $m=0$~\cite{Heusler:1996} 
\begin{equation} \label{eq:MaxCirc}
F(k,m) = 0 = \star F(k,m).    
\end{equation}
Meaning that the right-hand sides of \eqref{eq:MaxId} are in fact zero. Thus, the smooth functions being differentiated on the left-hand sides of \eqref{eq:MaxId} must be global constants. Given that these constant also vanish at the symmetry axis, the Frobenius integrability conditions follow, which constitute the electrovac circularity theorem.

In particular, for a stationary axisymmetric Kerr-Schild ansatz \eqref{eq:KSansatz}, necessarily built with the ellipsoidal flat seed metric \eqref{eq:FlatEllips} and the congruence \eqref{eq:lEllips}, the Frobenius integrability conditions turn into
\begin{equation}\label{eq:PDE(S)}
\begin{alignedat}{2}
0 &= \star(k\wedge m\wedge \rd k) &&= -\frac{2\sin\theta}{\Sigma}\partial_{\theta}\left(\Sigma S\right),\\
0 &= \star(k\wedge m\wedge \rd m) &&= \frac{2a\sin^3\theta}{\Sigma}\partial_{\theta}\left(\Sigma S\right).
\end{alignedat}
\end{equation}
From these latter, we are able to conclude that not all metric profiles are compatible with circularity by construction. Something similar occurs for the electric profile \eqref{eq:electric}, since the first of the electromagnetic circularity conditions \eqref{eq:MaxCirc} is identically satisfied, but the second turn out to be
\begin{equation}
    0=\star F(k,m) = \frac{\sin\theta}{\Sigma}\partial_{\theta}\left(\Sigma S_\text{e}\right).
\end{equation}
Hence, the electrovac circularity theorem restrict the stationary and axisymmetric Kerr-Schild profiles to
\begin{subequations}\label{eq:circSs}
\begin{align} 
    S(r,\theta) &= \frac{rM(r)}{\Sigma}, \label{eq:S}\\
    S_\text{e}(r,\theta) &= \frac{rQ(r)}{\Sigma}. \label{eq:Se}
\end{align}
\end{subequations}
For such profiles it is possible to find Boyer-Lindquist-type coordinates where the metric is manifestly circular \cite{Boyer:1966qh}. The remaining step is to solve the Einstein-Maxwell system that becomes straightforward in such coordinates; since circularity determines the explicit angular dependence \eqref{eq:circSs} of the profiles, the system becomes separable leading to simple ordinary differential equations. The result, in order of integration, is the following 
\begin{subequations}
\begin{align}
        Q(r)&=q,\label{eq:Q(r)}\\
        M(r)&= m-\frac{q^2}{2r},\label{eq:M(r)e}
\end{align}
\end{subequations}
where the integration constants are recognized as the electric charge and the mass of the now fully determined original electrically charged Kerr-Newman black hole \cite{Newman:1965my}.

%%%%%%%%%%%%%%%%%%%%%%%%%%%%%%%%%%%%%%%%%%%%%%%%%
\section{The Dyonic Extension} \label{sec:dyonic}
%%%%%%%%%%%%%%%%%%%%%%%%%%%%%%%%%%%%%%%%%%%%%%%%%

Our approach will consist in taking advantage of the purely electric derivation to construct the dyonic extension. We start by pointing out the following: the fact that the purely electric solution---expressed by ansatz \eqref{eq:electric} together with expressions \eqref{eq:Se} and \eqref{eq:Q(r)}---satisfies Maxwell equations is due to the null vector \eqref{eq:lEllips} obeys the geometric identity 
\begin{equation}\label{eq:dsd}
    \rd\star\rd\left(\frac{r}{\Sigma}l\right)=0.
\end{equation}
In other words, the two-form above is closed. Thus, by the Poincar\'e lemma, it follows that it is also locally exact and a one-form potential exists for it. It is straightforward to show that the most general of these potentials is conveniently written as
\begin{equation}\label{eq:exact}
    \star\,\rd\left(\frac{r}{\Sigma}l\right)=
    - \rd\left(\frac{\cos\theta}{\Sigma}l^{*}\right),
\end{equation}
in terms of the one-form
\begin{equation}\label{eq:l*Ellips}
    l^{*} = a \rd t - (r^2+a^2)\rd \phi+\frac{\Sigma}{\cos\theta}\rd f,
\end{equation}
where $f=f(x^\mu)$ is an arbitrary function. The above process is reminiscent of a  sort of Poincar\'e dualization \cite{Ortin:2004ms}, that is why we informally dub $l^{*}$ the ``dualized'' vector of the original null one $l$. 

We are now ready to make our proposal for the \emph{dyonic Kerr-Schild ansatz} in the electromagnetic context, consisting in supplementing metric \eqref{eq:KSansatz} with 
\begin{equation}\label{eq:dyonicAnsatz}
    A = -S_\text{e} \, l + S_\text{m} \, l^{*},
\end{equation}
that incorporates an independent magnetic contribution along the dualized vector to the classical purely electric ansatz \eqref{eq:electric}. In order to respect the stationarity and axisymmetry of the problem, we additionally assume the appropriate dependence on both profiles $S_\text{e}=S_\text{e}(r,\theta)$ and $S_\text{m}=S_\text{m}(r,\theta)$.

As can be anticipated, not all gauge potentials in this larger class respect the circularity conditions. Thus, our next step is to impose electromagnetic circularity \eqref{eq:MaxCirc} on the dyonic ansatz \eqref{eq:dyonicAnsatz}. The vanishing of the component $F(k,m)$ is automatically satisfied again, while, the other requirement reads
\begin{align}\label{eq:*F(k,m)}
0={}&\star F(k,m) = \frac{\sin\theta}{\Sigma}\Bigg[
\partial_\theta\left(\Sigma S_\text{e}\right) 
+\Delta\partial_r\left(\frac{\Sigma S_\text{m}}{\cos\theta}\right)\partial_\theta f \notag\\
&+\partial_\theta\left(\frac{\Sigma S_\text{m}}{\cos\theta}\right)\left[2\Sigma S \, l(f) -(r^2+a^2)\partial_r f\right]\Bigg],
\end{align}
where $\Delta = r^2+a^2-2\Sigma S$ and here $l$ is the contravariant version of the vector field \eqref{eq:lEllips} used as a derivation. Since the electromagnetic circularity must be satisfied for any function $f$ defining the class \eqref{eq:l*Ellips}, this requires the same behavior \eqref{eq:Se} for the electric profile while the magnetic one is restricted by 
\begin{equation}
       \partial_\theta\left( \frac{S_\text{m}\Sigma}{\cos\theta}\right)=0=\partial_r\left( \frac{S_\text{m}\Sigma}{\cos\theta}\right).
\end{equation}
These constraints fix the magnetic profile as
\begin{equation}\label{eq:Sm}
    S_\text{m} = \frac{p\cos\theta}{\Sigma},
\end{equation}
where $p$ is an integration constant afterwards identified as the magnetic charge. Using the above profile in \eqref{eq:dyonicAnsatz} it is now obvious that the contribution of the arbitrary function $f$ in the dualized vector \eqref{eq:l*Ellips} can be gauged out as
\begin{equation}
    A \mapsto A-\rd(p f),
\end{equation}
and harmoniously the whole class \eqref{eq:l*Ellips}  only reflects the electromagnetic gauge freedom. 

Considering that electromagnetic circularity is already verified, Frobenius integrability conditions \eqref{eq:PDE(S)} are again satisfied which restricts the metric profile another time as \eqref{eq:S}. Now we are able to move from the Kerr-Schild ansatz \eqref{eq:KSansatz} to write the metric in a more natural block-diagonal form by employing Boyer-Lindquist coordinates \cite{Boyer:1966qh}
\begin{align}\label{eq:BoyerLindquist}
    \tilde{t} &= t - \int\frac{2rM(r)}{\Delta}\rd r, &
    \tilde{\phi} &= \phi - \int \frac{2arM(r)}{(r^2+a^2)\Delta}\rd r,
\end{align}
where now 
\begin{equation}
    \Delta = r^2+a^2-2rM(r),
\end{equation}
becomes a function of $r$ only. It remains to solve Einstein-Maxwell equations in such coordinates for the two functions $Q(r)$ and $M(r)$. Maxwell's equations once more fix the function $Q(r)$ to a constant as in the purely electric case \eqref{eq:Q(r)}. Besides, there is a single independent Einstein equation 
\begin{equation}
    -\frac{\Sigma^2}{2r^2}\left(G^r_r - 2T^r_r\right) = \frac{\rd M}{\rd r}-\frac{q^2+p^2}{2r^2} = 0,
\end{equation}
and one ends up correcting \eqref{eq:M(r)e} to 
\begin{equation}\label{eq:M(r)d}
    M(r) = m -\frac{q^2+p^2}{2r}.
\end{equation}
This allows us to finally write the dyonic Kerr-Newman solution in its standard form \cite{Newman:1965my,Demianski:1966}
\begin{subequations}
    \begin{align}
    \rd s^2 ={}&-\frac{\Delta}{\Sigma}\!\left(\rd\tilde{t}-a\sin^2\theta \rd\tilde{\phi}\right)^2+\Sigma\!\left(\frac{\rd r^2}\Delta + \rd\theta^2\right)\notag\\
    &+\frac{\sin^2\theta}{\Sigma}\!\left(a\rd\tilde{t}-(r^2+a^2)\rd\tilde{\phi}\right)^2, \\
    A={}& -\frac{qr}{\Sigma}\left(\rd\tilde{t}-a\sin^2\theta \rd\tilde{\phi}\right)\notag\\
    &+\frac{p\cos\theta}{\Sigma}\left(a\rd\tilde{t}-(r^2+a^2)\rd\tilde{\phi}\right).
    \end{align}
\end{subequations}

%%%%%%%%%%%%%%%%%%%%%%%%%%%%%%%%%%%%%%%%%%%%
\section{Adding a Cosmological Constant} \label{sec:cosmological}
%%%%%%%%%%%%%%%%%%%%%%%%%%%%%%%%%%%%%%%%%%%%

In this section, we demonstrate how our approach can seamlessly extend to include a cosmological constant. By incorporating the cosmological constant into Einstein's equations, we observe that its presence does not disrupt the foundational structure of the method. This robustness highlights the adaptability of the approach and ensures its applicability to spacetimes that are asymptotically de Sitter or anti-de Sitter. A similar strategy to Ref.~\cite{Ayon-Beato:2015nvz} was used in \cite{Ayon-Beato:2015qtt} to derive asymptotically (A)dS black holes in bigravity, where at least one of the metrics is electrically charged. 

Perhaps, the most striking difference that should be emphasized is that despite that the Kerr theorem \cite{Debney:1969zz,Cox:1976,Stephani:2003tm} can be straightforwardly extended to (A)dS exploiting conformal flatness, there exists no analogue yet of the stationary and axisymmetric version proved for Minkowski spacetime in \cite{Ayon-Beato:2015nvz}. In fact, it has proven to be a difficult problem, see \cite{Higuita-Borja:2018} for partial advances. Nevertheless, an explicit shear-free and geodesic null congruence on (A)dS that happens to be additionally stationary and axisymmetric was provided by Carter long ago in \cite{Carter:1972,*Carter:2009nex}. In the notation of \cite{Gibbons:2004uw},  it is given by
\begin{equation}\label{eq:lCarter}
l=\frac{\Delta_\theta}{\Xi}\rd t
-\frac{a\sin^2\theta}{\Xi}\rd\phi
+\frac{\Sigma}{(1-\lambda r^2)(r^2+a^2)}\rd r,
\end{equation}
where, in the related coordinates, the (A)dS metric is also manifestly stationary and axisymmetric 
\begin{align}\label{eq:(A)dSCarter}
\rd s_{0}^2= & -\frac{(1-\lambda r^2)\Delta_{\theta}}{\Xi}\rd t^2
+\frac{(r^2+a^2)\sin^2{\theta}}{\Xi}\rd\phi^2\nonumber\\
& +\frac{\Sigma}{(1-\lambda r^2)(r^2+a^2)}\rd r^2
+\frac{\Sigma}{\Delta_{\theta}}\rd\theta^2,
\end{align}
being $\Lambda=3\lambda$ the involved cosmological constant. We also use the notation
\begin{equation}
    \Delta_\theta=1+\lambda a^2\cos^2\theta,
\end{equation}
and the factor $\Xi=1+\lambda a^2$ is introduced to avoid a conical singularity as usual. We still use the definition $\Sigma=r^2+a^2\cos^2\theta$. As previously discussed, the uniqueness of the Carter congruence \eqref{eq:lCarter} cannot be ensured, which however does not rule out that it is a good starting point for studying stationary and axisymmetric Kerr-Schild transformations \eqref{eq:KSansatz} from (A)dS \eqref{eq:(A)dSCarter}.

As has been shown in~\cite{Ayon-Beato:2015qtt}, for a  
Kerr-Schild ansatz like the previous one the electric solution given by \eqref{eq:Se} with \eqref{eq:Q(r)} follows from the proportionality ansatz \eqref{eq:electric}. Accordingly, the tangent vector $l$ to the Carter congruence \eqref{eq:lCarter} also satisfies the closedness condition \eqref{eq:dsd} and its dualized vector is again defined from the exactness condition \eqref{eq:exact} giving now
\begin{equation}
    l^{*} = \frac{1-\lambda r^2}{\Xi}a\rd t - \frac{r^2+a^2}{\Xi}\rd \phi+\frac{\Sigma}{\cos\theta}\rd f,
\end{equation}
where as before, $f$ is an arbitrary function. All the above ingredients allow to improve the purely electric ansatz \eqref{eq:electric} to the dyonic Kerr-Schild ansatz following once more the proposal \eqref{eq:dyonicAnsatz} in the (A)dS context. Notice that consistently, the value $\lambda=0$ reproduces the dyonic Kerr-Schild ansatz from the previous section.

The presence of the cosmological constant has no effect on the electrovac circularity theorem since the identities \eqref{eq:MaxId} are still satisfied. Examining now the electromagnetic circularity \eqref{eq:MaxCirc} for the (A)dS dyonic ansatz provides a condition similar to \eqref{eq:*F(k,m)}. This again ensures  that the magnetic profile is completely fixed as \eqref{eq:Sm} in terms of the magnetic charge, while the electric one acquires the explicit angular dependence \eqref{eq:Se}. As a result, the circularity conditions are given by
\begin{equation}\label{eq:PDE(S)Lamda}
\begin{alignedat}{2}
0 &= \star(k\wedge m\wedge \rd k) &&= -\frac{2\sin\theta\Delta_{\theta}^2}{\Xi^2\Sigma}\partial_{\theta}\left(\Sigma S\right),\\
0 &= \star(k\wedge m\wedge \rd m) &&= \frac{2a\sin^3\theta\Delta_{\theta}}{\Xi^2\Sigma}\partial_{\theta}\left(\Sigma S\right).
\end{alignedat}
\end{equation}
Thus, in order for the studied stationary-axisymmetric Kerr-Schild-(A)dS ansatz to be circular, the metric profile must be restricted just like in absence of the cosmological constant \eqref{eq:S}. This warrants the existence of metric block-diagonal Boyer-Lindquist coordinates~\cite{Boyer:1966qh}
\begin{equation}\label{eq:BoyerLindquist(A)dS}
\begin{aligned}
    \tilde{t} &= \Xi t - \int\frac{2\Xi rM(r)}{(1-\lambda r^2)\Delta_r}\rd r, \\
    \tilde{\phi} &= \phi+\lambda at - \int \frac{2a\Xi rM(r)}{(r^2+a^2)\Delta_r}\rd r,
\end{aligned}
\end{equation}
where we have defined the radial function as
\begin{equation}
    \Delta_r = (r^2+a^2)(1-\lambda r^2)-2rM(r).
\end{equation}
Finally, even in presence of the cosmological constant the remaining independent radial functions are easily determined from the Einstein-Maxwell equations as in the previous section, i.e., settling the electric charge and the mass from \eqref{eq:Q(r)} and \eqref{eq:M(r)d}, respectively. Which brings us to the dyonic Kerr-Newman-(A)dS black hole solution~\cite{Carter:1968ks,Carter:1972,*Carter:2009nex,Plebanski:1975xfb}
\begin{subequations}
    \begin{align}
    \rd s^2 ={}& -\frac{\Delta_r}{\Xi^2\Sigma}\!\left(\rd\tilde{t}
    -a\sin^2\theta\rd\tilde{\phi}\right)^2
    +\Sigma\!\left(\frac{\rd r^2}{\Delta_r}
    +\frac{\rd\theta^2}{\Delta_{\theta}}\right)\notag\\
    &+\frac{\Delta_\theta\sin^2\theta}{\Xi^2\Sigma}\!\left(a\rd \tilde{t} - (r^2+a^2)\rd \tilde{\phi}\right)^2, \\
    A ={}& -\frac{qr}{\Xi\,\Sigma}\left(\rd\tilde{t}
    -a\sin^2\theta\rd\tilde{\phi}\right)\notag\\
    & + \frac{p\cos\theta}{\Xi\,\Sigma}\left(a\rd \tilde{t} - (r^2+a^2)\rd \tilde{\phi}\right) .
    \end{align}
\end{subequations}

%%%%%%%%%%%%%%%%%%%%%%%%%%%%%%%%%%%%%%%%%%%%
\section{Conclusions}
%%%%%%%%%%%%%%%%%%%%%%%%%%%%%%%%%%%%%%%%%%%%

This work presents a new, geometrically grounded method for deriving dyonically charged black holes within the Kerr–Schild framework. Traditionally, the inclusion of a magnetic charge on top of an electrically charged solution is achieved via an external electromagnetic duality rotation, independent of the Kerr–Schild structure. In contrast, our approach integrates both electric and magnetic sectors directly into the ansatz itself. Central to this construction is the shear-free and geodesic null congruence underlying the Kerr–Schild ansatz—reminiscent of the one defining the vacuum Kerr black hole—which naturally leads to a second one-form encoding the magnetic contribution through a procedure inspired by the Poincar\'e lemma. This Poincar\'e dualized one-form is not externally imposed, but arises intrinsically from a geometric closedness condition satisfied by the stationary axisymmetric congruence itself. Ultimately, allowing the electric and magnetic contributions of the Maxwell field to be treated on equal footing in the context of black holes. All the involved scalar Kerr-Schild profiles are uniquely constrained by the circularity conditions, ensuring consistency of the resulting dyonic configuration. We have also shown that the method extends seamlessly to incorporate a cosmological constant, demonstrating the robustness of the formalism in (A)dS backgrounds.

A natural extension of this work would be to test the method in more elaborate scenarios involving additional fields or nontrivial matter couplings. A particularly compelling example involves the dyonic rotating black hole of Ref.~\cite{Clement:2004yr}, found in Einstein–Maxwell theory supplemented with a fixed dilaton coupling. That solution, constructed via dimensional reduction from five-dimensional vacuum gravity and a subsequent duality transformation, presents an ideal testing ground for assessing whether our geometric procedure can accommodate scalar fields and reproduce the dyonic structure without invoking external dualization.

Returning to the vacuum case, the null congruence \eqref{eq:lEllips} employed in the construction was uniquely selected by the refined Kerr theorem, formulated in its stationary and axisymmetric version \cite{Ayon-Beato:2015nvz}. It was shown there that this congruence uniquely depends on the angular momentum parameter, since the remaining conserved quantities in involution can be fixed by exploiting the residual gauge freedoms and symmetries of the problem. This provided a solid geometric basis, and ultimately the uniqueness, of the resulting stationary axisymmetric Kerr-Schild solutions. In contrast, no analogous uniqueness result has been established in the presence of a cosmological constant for the stationary axisymmetric Carter congruence \eqref{eq:lCarter}. Part of the problem in (A)dS is that translational symmetries are replaced by quasi-translations, which no longer commute. As a result, once the (A)dS spacetime is written in a manifestly stationary and axisymmetric coordinate system, no further symmetries remain manifest. One can resort to geodesic hidden symmetries and their Killing-tensor constructed conserved quantities as the square of angular momentum, or even use conformal Killing tensors, but this does not prevent the integration from being highly involved. Consequently, one expects the appearance of additional kinematical parameters, beyond the angular momentum, characterizing the class of stationary-axisymmetric, shear-free, and geodesic null congruences in (A)dS. Whether this extra parameters can be eliminated through coordinate or gauge transformations—thus singling out the Carter congruence \eqref{eq:lCarter}—remains an open question. Establishing such a uniqueness result would not only help clarify the geometric role of the Carter congruence but also implies the uniqueness of the Kerr-Schild associated (A)dS rotating solutions. We leave this question for future investigation (see partial advances in \cite{Higuita-Borja:2018}).

A promising direction for future work would be to explore whether the geometric framework developed here, and based on the proposed dualization of the Kerr–Schild congruence, can be extended to construct dyonic rotating solutions also in nonlinear theories. Recent results have established the existence of nonlinearly charged rotating black hole solutions \cite{Garcia-Diaz:2021bao, DiazGarcia:2022jpc}, and the underlying theory supporting them was subsequently given in explicit form \cite{Ayon-Beato:2022dwg}. The vector potentials of such configurations exactly exhibit a dyonic superposition similar to \eqref{eq:dyonicAnsatz}, concretely see Eq.~(19) in \cite{Ayon-Beato:2022dwg}. However, the involved superposed one-forms are no longer dualized through the exactness condition \eqref{eq:exact} rooted in the linear Maxwell equations, and that the related nonlinear purely electric solutions cease to satisfy. This raises a natural and challenging question: can a second one-form spanning the magnetic contribution, still be generated geometrically via a dualization mechanism in such nonlinear settings? In a similar way, new developments in the ModMax theory \cite{Bandos:2020jsw} have shown that certain spacetimes such as the Schwarzschild–Melvin–Bonnor solution and the accelerating C-metric can be cast in Kerr–Schild and double Kerr–Schild form, respectively \cite{Barrientos:2024umq}. Also, a novel Kerr-Schild ansatz was proposed in Ref.~\cite{Hassaine:2024mfs}, where the full black hole geometry is expressed as a linear in mass perturbation of the associated extremal black hole, which now serves as the seed metric. Applying our approach to all the above configurations offer new perspectives on how (non)linearity and duality are encoded geometrically, and could point toward a more universal formulation of dyonic configurations within the Kerr–Schild paradigm.

Finally, it would be interesting to explore in future work how our method could be extended to higher-dimensional settings involving extended objects such as $p$-branes. In supergravity and string theory, many charged black brane solutions are known which typically carry either electric or magnetic charges under various gauge fields, see e.g.~\cite{Green:1996vh}. Extending our geometric framework to accommodate both electric and magnetic components simultaneously, that is to construct genuine dyonic p-branes, could provide new insights into the interplay between geometry and duality in higher dimensions. This would involve generalizing the underlying Kerr–Schild structure and investigating whether a similar mechanism, possibly involving higher-rank forms and generalized congruences, could encode the dual charges in a unified geometric way.

~

%%%%%%%%%%%%%%%%%%%%%%%%
\begin{acknowledgments}
%%%%%%%%%%%%%%%%%%%%%%%%
D.F.-A.\ is supported by SECIHTI through a postdoctoral research grant. M.H.\ would like to thank the University of Paris Saclay for their kind hospitality where part of this work has been realized. D.F.H.-B.\ is also grateful to SNII from SECIHTI.
\end{acknowledgments}

\end{document}